\documentclass[fleqn,usenatbib]{mnras}

\usepackage{newtxtext,newtxmath}
\usepackage[T1]{fontenc}
\DeclareRobustCommand{\VAN}[3]{#2}
\let\VANthebibliography\thebibliography
\def\thebibliography{\DeclareRobustCommand{\VAN}[3]{##3}\VANthebibliography}

\usepackage{amstext}
\usepackage{url}
\usepackage[]{times,refname,amsmath,bm}
\bibpunct{(}{)}{;}{a}{}{,}
\usepackage{tabularx}
\usepackage{graphicx,epsfig,color,latexsym}
\usepackage{orcidlink}
\usepackage[english]{babel}

\newcommand{\bea}{\begin{eqnarray}}
\newcommand{\eea}{\end{eqnarray}}
\newcommand{\be}{\begin{equation}}
\newcommand{\ee}{\end{equation}}

\usepackage{ulem}

\title[Plasma lensing in QSOs]{An observational test of the plasma lensing effect using QSOs with and without MgII absorption}
\author[Er, Shu \& Liu]{Xinzhong Er$^{1}$\orcidlink{0000-0002-8700-3671}, 
Yiping Shu$^{2}$\orcidlink{0000-0002-9063-698X},  Chenxu Liu$^{3}$\orcidlink{0000-0001-5561-2010}\\
$^1$Tianjin Astrophysical Center, Tianjin Normal University, Tianjin, 300387, P.R. China, phioen@163.com\\
$^2$Purple Mountain Observatory, Chinese Academy of Sciences, Nanjing, 210008, P.R. China, yiping.shu@pmo.ac.cn\\
$^3$South-Western Institute for Astronomy Research (SWIFAR), Yunnan University, Kunming, Yunnan 650500, P.R. China
}
\pubyear{2026}

\begin{document}
\label{firstpage}
\pagerange{\pageref{firstpage}--\pageref{lastpage}}
\maketitle

% Abstract of the paper
\begin{abstract}
Radio wave propagation can be perturbed by compact ionized gas clumps through plasma lensing, which induces frequency dependent magnification and may distort the observed number counts of background sources. 
The quasar (QSO) number densities are a powerful probe for understanding the effects of intervening material. Absorption lines in QSO spectra reveal the presence of interstellar and intergalactic gas, which can change observed fluxes through dust extinction and plasma lensing. By combining observations from radio (VLASS), infrared (WISE), and optical bands (DESI), we assembled a sample of QSOs: $\sim4000$ sources with MgII absorbers, and $\sim12,000$ non-absorbers. 
In the radio band, the MgII sample shows a moderate excess at the bright end of the flux distribution, which is broadly consistent with plasma lensing predications. In the optical, the MgII sample turns over at higher $g$-band fluxes and exhibits a steeper decline at the faint end than the non-MgII sample. Control samples were constructed by matching in redshift, infrared (W1), and optical ($g$) luminosities. In these comparisons, the radio excess becomes less prominent, suggesting that the apparent magnification may not be robust evidence for plasma lensing. Nevertheless, a weak contribution cannot be ruled out, especially given residual excess observed at the bright end relative to the non-MgII sample. Dust extinction along the line-of-sight remains a plausible alternative. Regardless of the dominant mechanism, the multi-wavelength differences offer a valuable probe of the physical state of the intervening medium. 
\end{abstract}
%207
\begin{keywords}
quasars: general -- gravitational lensing: strong -- intergalactic medium 
\end{keywords}

%%%%%%%%%%%%%%%%% BODY OF PAPER 
\section{Introduction}
Quasi-Stellar Objects (QSOs) are luminous and compact astrophysical sources, with energy outputs exceeding those of entire host galaxies, and typically exhibit bolometric luminosities in the range $10^{45}-10^{48}$ erg\,s$^{-1}$. Since their discovery \citep[e.g.][]{1963Natur.197.1040S,1963ApJ...138...30M}, millions of QSOs have been identified \citep[e.g.][]{2022ApJ...935..132L,2023OJAp....6E..49F}. Their broadband spectral energy distributions (SEDs), spanning from radio to high-energy wavelengths, arise from multiple physical processes: synchrotron radiation in the radio bands is detected from a fraction of optically selected sources \citep[e.g.][]{2013ApJ...768...37C}, thermal emission from dust dominates in the infrared \citep[e.g.][]{1994ApJS...95....1E}, optical, ultraviolet, and X-ray emission originate from the accretion processes near the central engine. These extreme luminosities make QSOs powerful probes of galaxy evolution and cosmology at various scales, including active galactic nucleus (AGN) feedback mechanisms and the growth of supermassive black holes \citep[e.g.][]{2013ARA&A..51..511K,2023ARA&A..61..373F}. They also serve as bright background sources for studying the properties of the interstellar and intergalactic medium, as well as the the history of cosmic reionization \citep[e.g.][]{2007ApJ...655..735H}.

The spectra of QSOs serve as powerful backlights for probing the intervening universe. Absorption lines imprinted on these spectra provide a unique method to study the physical conditions, chemical composition, and spatial distribution of baryonic matter from nearby environment of galaxies to the diffuse intergalactic medium \citep[e.g.][]{2001AJ....122..549V,2005pgqa.conf...24C,2013ApJ...770..130Z,2018ApJ...868..142R,2018ApJ...866...36L}. In particular, the MgII doublet is a powerful and widely adopted diagnostic, owing to a combination of practical and physical advantages. Observationally, its distinctive doublet structure is easy to identify in spectra over a broad redshift range, making it an efficient tracer for large statistical studies \citep[e.g.][]{2013ApJ...770..130Z,2023AJ....166...99N}. Physically, MgII traces cool, metal-enriched gas, making it a probe for the multiphase foreground clouds \citep[e.g.][]{2000ApJS..130...91C,2005pgqa.conf...24C}.
The origins of these absorbers are diverse, arising in the interstellar media of foreground galaxies, extended galactic halos, and within the host galaxies of the QSOs themselves \citep[e.g.][]{2008ApJ...683...55C,2008MNRAS.388..227W}. QSOs exhibiting MgII absorption are observed to be systematically redder, suggesting the presence of associated dust of host galaxies \citep[e.g.][]{2023MNRAS.525.5575F}. 

In addition to absorption by intervening material, QSOs exhibit a diversity of intrinsic spectral properties. The majority display a characteristic blue power-law continuum in the UV and optical bands, consistent with thermal emission from an optically thick accretion disc. The discovery of a population with markedly redder continua has prompted debate regarding their physical relationship to the blue population. This reddening is frequently attributed to dust extinction of the intrinsic disk emission \citep[e.g.][]{2013MNRAS.432.2150R}. The underlying nature of these red QSOs remains unclear; they may represent a distinct evolutionary phase linking obscured star formation to AGN feedback through gas inflows and outflows, or they could be an artifact of viewing angle orientation. Support for an evolutionary scenario has been identified within radio selected samples \citep[e.g.][]{2019MNRAS.488.3109K}. The complex radio emission observed in radio-quiet QSOs suggests multiple mechanisms may also be at play \citep[e.g.][]{2024MNRAS.529.1995P}. Notably, recent studies report correlations between dust reddening and both radio detection fraction \citep[e.g.][]{2023MNRAS.525.5575F,2024A&A...691A.191C,2025MNRAS.537.2003F} and the detection rate of [OIII] outflows \citep{2025MNRAS.536.1166E}. 

Beyond absorption and reddening, the propagation of radio emission from QSOs through ionized gas can give rise to refractive phenomena, i.e. plasma lensing \citep[e.g.][]{2016ApJ...817..176T,2017ApJ...842...35C,2018MNRAS.475..867E,2019MNRAS.486.2809G}. Conceptually analogous to gravitational lensing \citep[e.g.][]{2022iglp.book.....M,2020arXiv200616263W}, plasma lensing was initially proposed to explain the anomalous flux variations in the radio light curves of QSOs, phenomena known as Extreme Scattering Events \citep{1987Natur.328..324R,CleggFL1998}. In this process, localized over-densities of ionized gas along the line of sight act as refractive diverging lenses, deflecting radio waves via interactions between low energy photons and free electrons. Because the refractive index of a plasma depends inversely on the square of the observing frequency, for typical interstellar or intergalactic plasma densities, observable signatures of plasma lensing are generally confined to low frequency radio bands, while being negligible at infrared, optical, or higher frequencies. 
Plasma lensing can shift the apparent positions of background sources, though such astrometric effects are typically at the milli-arcsec level making them difficult to detect. More importantly, it can induce frequency dependent magnification and demagnification \citep[e.g.][]{1987Natur.328..324R,2018MNRAS.475..867E,2024MNRAS.531.4155C}, potentially distorting the observed spectral energy distribution at low radio frequencies. Analogous to gravitational lensing, which systematically alters the observed luminosity function by boosting bright-end number counts \citep[e.g.,][]{1996MNRAS.283.1340B, 2010MNRAS.406.2352L}, plasma lensing is also expected to modify source number counts. Since plasma lensing can induce both magnification and demagnification effects, deviations may appear at both the bright and faint ends of the radio number count distribution \citep{2022MNRAS.510..197E}. This signature offers a potential probe of small scale structures in intervening ionized gas along the line of sight. Recent studies further suggest that plasma structures may influence the observed spatial distribution of fast radio bursts (FRBs). \citet{2025arXiv250906721S} report a notable absence of FRB detections toward a region coincident with a known plasma rich structure in the Milky Way, hinting at strong scattering or diverging by foreground ionized gas.

Due to the fact that plasma lensing is strongly frequency dependent, i.e. the deflection angle scales as $\lambda^2$, its effects are significant only at low radio frequencies. This motivates a multi-wavelength comparison of QSO number counts across the radio, infrared, and optical bands. In this study, we use the detection of MgII absorption lines as an indicator of the existence of foreground gas clumps. By comparing the number counts of QSOs with or without MgII absorbers, we aim to identify potential imprints of plasma lensing or other gas-related effects on the observed source populations. However, we emphasize that the presence of absorption lines is neither a sufficient nor a necessary condition for observable plasma lensing. Gas clumps located in close proximity to the background source exhibit low lensing efficiency due to the geometric leverage. Conversely, highly ionized, low-density gas that does not produce detectable absorption lines can still generate lensing effects.
The observational data used in this study are described in Sec.\,\ref{sec:data}, and the corresponding flux distributions and number count are presented in Sec.\,\ref{sec:lum-func}.

%%%%%%%%%%%%%%%%%%%%%%%%%%%%%%%%%%%

\section{data}
\label{sec:data}
\begin{figure*}
\centering
\includegraphics[width=18cm]{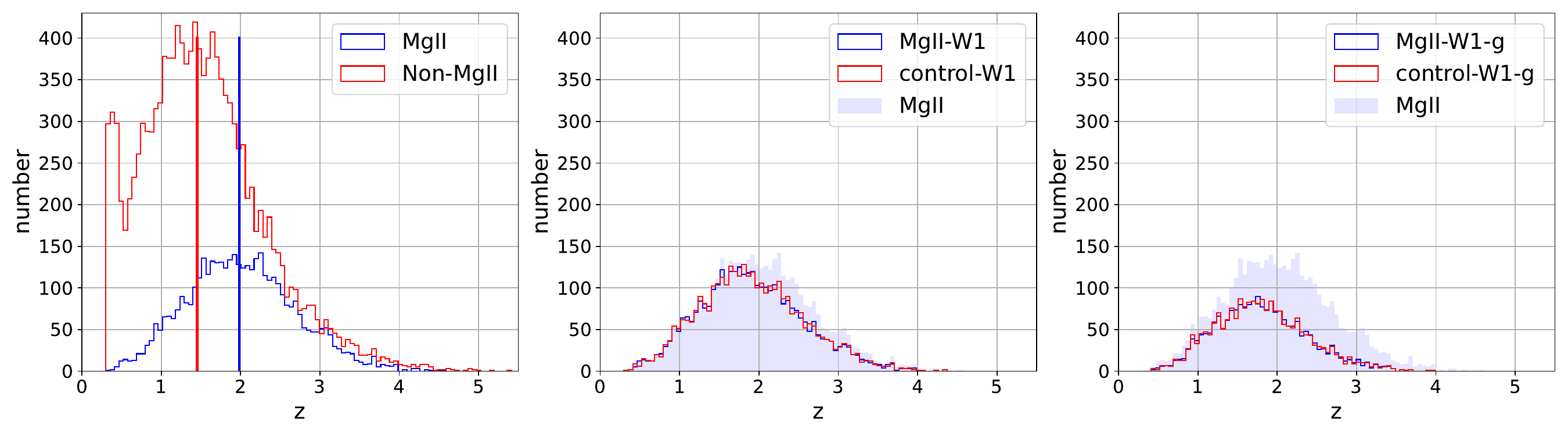}
\caption{Redshift distributions of the QSO samples used in this study. The blue histograms in the three panels show sources contains MgII absorption lines; and the red histograms show non-MgII (control) samples. Vertical lines in the left panel present the median redshifts of the initial MgII and non-MgII samples. In the middle and right panels, the shaded histogram represents the initial MgII sample before matching. }
\label{fig:z-dist}
\end{figure*}
\begin{figure*}
\centering
\includegraphics[width=18cm]{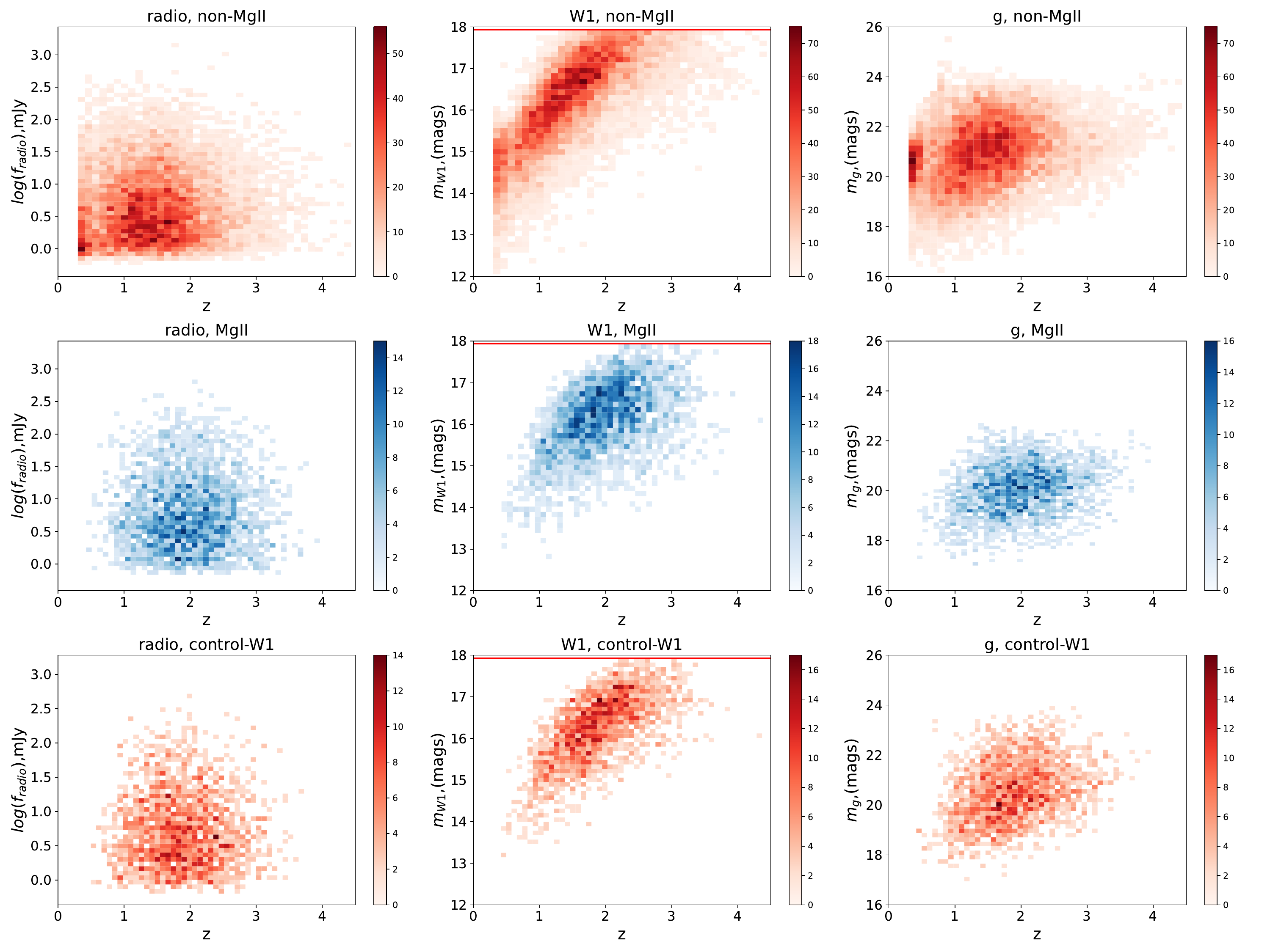}
\caption{The heat map of redshift vs flux in radio (left), redshift vs infrared Vega-magnitude (middle) and redshift vs optical $g$-band AB-magnitude (right). From top to bottom are non-MgII, MgII, and control-W1 samples respectively. The horizontal red lines in the middle panels show the cut in W1 magnitude (17.93). The colour bars represent the numbers of sources in each bin.}
\label{fig:z-2d-dist}
\end{figure*}

We use the data from the Dark Energy Spectroscopic Instrument \citep[DESI,][]{2016arXiv161100036D,2024AJ....167...62D} Data Release 1 \citep[DR1,][]{2025arXiv250314745D}, which contains $\sim$19 million objects. Radio counterparts are drawn from the Very Large Array Sky Survey \citep[VLASS,][]{2020PASP..132c5001L}, which covers the sky at 2-4 GHz (centred at 3 GHz) with an angular resolution of approximately $2''.5$, and a typical single epoch sensitivity of $70\mu Jy$, yielding over 2.6 million detected sources \citep{2021ApJS..255...30G}. Both samples are likely incomplete at the faint end, limiting our ability to probe potential demagnification effects in the faint end number counts. We cross match DESI and VLASS catalogues using a positional tolerance of $0.5$ arcsec, consistent with the positional accuracy of the VLASS catalog \citep[$\sim 0.5$ arcsec,][]{2021ApJS..255...30G}. The matching yields 73015 sources with detections in both radio and optical bands. For sources with multiple detections separated by less than $0.1$ arcsec, we retain only one entry to avoid duplication \citep{2023AJ....166...99N}. Since QSOs exhibit broad band emission, and MgII absorption can traces foreground gas clump, we restrict the sample to objects spectroscopically classified as QSOs \citep{2023ApJ...944..107C}, and obtain 20233 QSOs as our initial sample. To assess potential misidentifications, particularly between compact optical nuclei and extended radio lobes, we randomly select a few hundred sources for visual inspection. No significant mismatches attributable to faint core with bright lobes in radio images were found. Sources at redshift $z<0.3$ are excluded, as MgII absorption lines fall outside the DESI spectral coverage at these redshifts \citep{2023ApJ...944..107C}. %19013 sources remains. 
Infrared photometry is obtained from the {\it Wide-field Infrared Survey Explorer} \citep[WISE][]{2010AJ....140.1868W}, specifically the $3.4\mu$m (W1) and $4.6\mu$m (W2) bands. We apply a magnitude cut of 17.93 in the W1 band (Vega magnitude, corresponding to a flux limit of 5.6 nanomaggy), to ensure a $50\%$ completeness of the sample \citep{2019ApJS..240...30S}. 
%and W2 > 9 nanomaggy, corresponding to magnitudes 16.72, 
The final sample is cross matched with the DESI MgII absorber catalogue \citep{2023AJ....166...99N}, yielding two samples: 4242 MgII absorbers and 13155 non absorbers, labelled as MgII and non-MgII respectively (Table\,\ref{tab:data}). Approximately $24\%$ of the final sample exhibits MgII absorption. The purity of the MgII absorbers catalogue is high, reaching $99.1\%$. However, the completeness is lower, dropping to $\sim 50\%$ at low redshifts due to a reduced throughput at the blue end of the DESI instrument \citep{2022AJ....164..207D}. 
%This fraction is higher than that reported in \cite{2025AJ....170...16N}. This enhancement likely arises from our joint selection criteria based on redshift and W1 flux, which preferentially includes high-z QSOs that have a greater probability of intersecting foreground gas clumps along the line of sight.

The redshift distributions of the MgII and non-MgII samples are shown as red and blue histograms in the left panel of Fig.\,\ref{fig:z-dist}. The two samples exhibit markedly distinct behaviours: the MgII sample peaks at $z\sim2$, whereas the non-MgII sample shows a primary peak at $z\sim 1.5$ and a secondary peak at $z\sim 0.4$. Investigating the lower redshift peak, we find that most of the sources were initially targeted by DESI as bright galaxies, which are selected to be low redshift, but were later spectroscopically confirmed as QSOs \citep{2023AJ....165..253H}.
The higher redshift peak of the MgII sample is consistent with the expectation that longer sight lines through the intergalactic medium increase the probability of intersecting an MgII absorbing cloud \citep[e.g.][]{2023AJ....166...99N}. Additionally, as pointed out, the MgII sample may suffer from lower completeness at low redshifts \citep{2023AJ....166...99N}.
To ensure a fair comparison between absorbers and non-absorbers, we constructed control samples using a one-to-one pairing between the MgII and non-MgII samples. We use the W1 band as a proxy for the intrinsic AGN luminosity, as it is minimally affected by dust extinction, while the optical $g$-band is sensitive to both accretion disk emission and selection effects associated with dust reddening and target selection. By constructing control samples, we aim to disentangle intrinsic luminosity effects from observational selection and dust-related biases in the radio number counts. For each source in the MgII sample, we identified a counterpart in the non-MgII sample based on redshift ($|\Delta z| < 0.01$) and luminosity constraints. We implemented two matching schemes: 
\begin{itemize}
\item 
Infrared only matching: sources were paired to match the W1-band luminosity within $\Delta logL_{W1}<0.05$ dex, yielding the control-W1 sample.
\item 
Joint optical-infrared matching: both $L_g$ and $L_{W1}$ were required to satisfy $\Delta logL<0.05$ dex simultaneously, resulting in the control-W1-g sample. 
\end{itemize}
Sources in the MgII sample without a suitable match in the non-MgII sample under a given scheme were excluded from the analysis. Consequently, in each comparison, the MgII and corresponding control samples contain identical number of objects. The corresponding MgII samples are named as MgII-W1 and MgII-W1-g sample.
The joint matching yields a significantly smaller sample size, primarily due to the fact that the optical and infrared luminosity distributions of the two parent populations are not fully aligned. To assess robustness, the entire matching procedure was repeated multiple times using a randomised selection and consistent results were obtained. The final sample sizes are listed in Table\,\ref{tab:data}. The redshift distributions of the control-W1 and control-W1-g samples are shown in Fig.\,\ref{fig:z-dist} as well. The reduction in source count from the initial MgII sample to the control-W1-g sample occurs primarily at high redshifts, e.g., $z>1.5$.

\begin{table}
\centering
\begin{tabular}{c|c|c|c}
    &Initial  &$z>0.3$ &$W1<17.93$ \\
 \hline
 \hline
 MgII absorption &4371 &4369 &4242 \\
% m-MgII &    &   &1703  &1551 \\
 \hline
 non-MgII absorption &15862 &14644 &13155  \\
 \hline
 % \hline
 % &control-W1 &control-g    &control-g-W1\\
 %  &3631  &3566  &1722\\
\end{tabular}
\caption{The numbers of QSOs from DESI DR1 at each stage of selection. The number of sources in control samples is 3477 (control-W1), and 2127 (control-W1-g). The MgII samples are matched in number correspondingly. } 
\label{tab:data}
\end{table}

\section{Behaviours in multiple bands}
\label{sec:lum-func}
Fig.\,\ref{fig:z-2d-dist} presents the two-dimensional distributions of sources in the radio, WISE W1, and optical $g$-band for the three primary samples: non-MgII, MgII, and the control-W1 sample. A clear anti-correlation between W1 magnitude and redshift is evident in all samples. In contrast, the radio flux shows no discernible redshift dependence. We further examine fluxes in other bands, infrared (W2) and optical ($r$ and $z$), and confirm a similar anti-correlation in W2. 
We attribute this redshift dependent trend in the infrared bands to a selection effect inherent to the DESI target selection algorithm. The algorithm employs a combination of optical and optical+infrared colour cuts \citep{2023ApJ...944..107C}, which likely biases the sample against sources with low infrared flux.

QSOs with absorption lines are evidence of the presence of foreground gas clumps along the line of sight. Thus the photometry of these QSOs are likely affected by such intervening material, such as dust extinction or plasma lensing. The effects of plasma lensing are most pronounced at low radio frequencies, where they can alter the observed flux density \citep[e.g.][]{2022MNRAS.510..197E}. We therefore hypothesize that the radio luminosity function of the MgII sample may be modified by this lensing effects. At infrared and higher frequencies, plasma lensing is negligible \citep[][]{2016ApJ...817..176T}, and thus unlikely to influence optical or IR photometry. In contrast, QSOs without detected MgII absorption are less likely to be intersected by such foreground structures. 
Dust obscuration presents another distinct effect: it can suppress optical emission while enhancing IR and radio flux through re-radiation \citep[e.g.][]{2019MNRAS.488.3109K}. To investigate the potential influence of these effects, we examine correlations between the radio flux and other photometric bands. Fig.\,\ref{fig:color2fr} shows the distribution of sources in the plane of radio flux versus the optical-infrared colour index ($g-W1$). No significant correlation is observed in either the MgII or non-MgII samples. However, the non-MgII sample contains a large fraction of redder sources (larger $g-W1$ values). In Fig.\,\ref{fig:corr-w1-radio}, we display the relation between infrared (W1) and radio flux. A weak positive correlation is evident in the MgII sample, as well as in the control-W1 samples, but is absent in the non-MgII sample. A similar trend is observed in the comparison between optical $g$-band and radio flux (Fig.\,\ref{fig:corr-g-radio}).

\begin{figure}
\includegraphics[width=8cm]{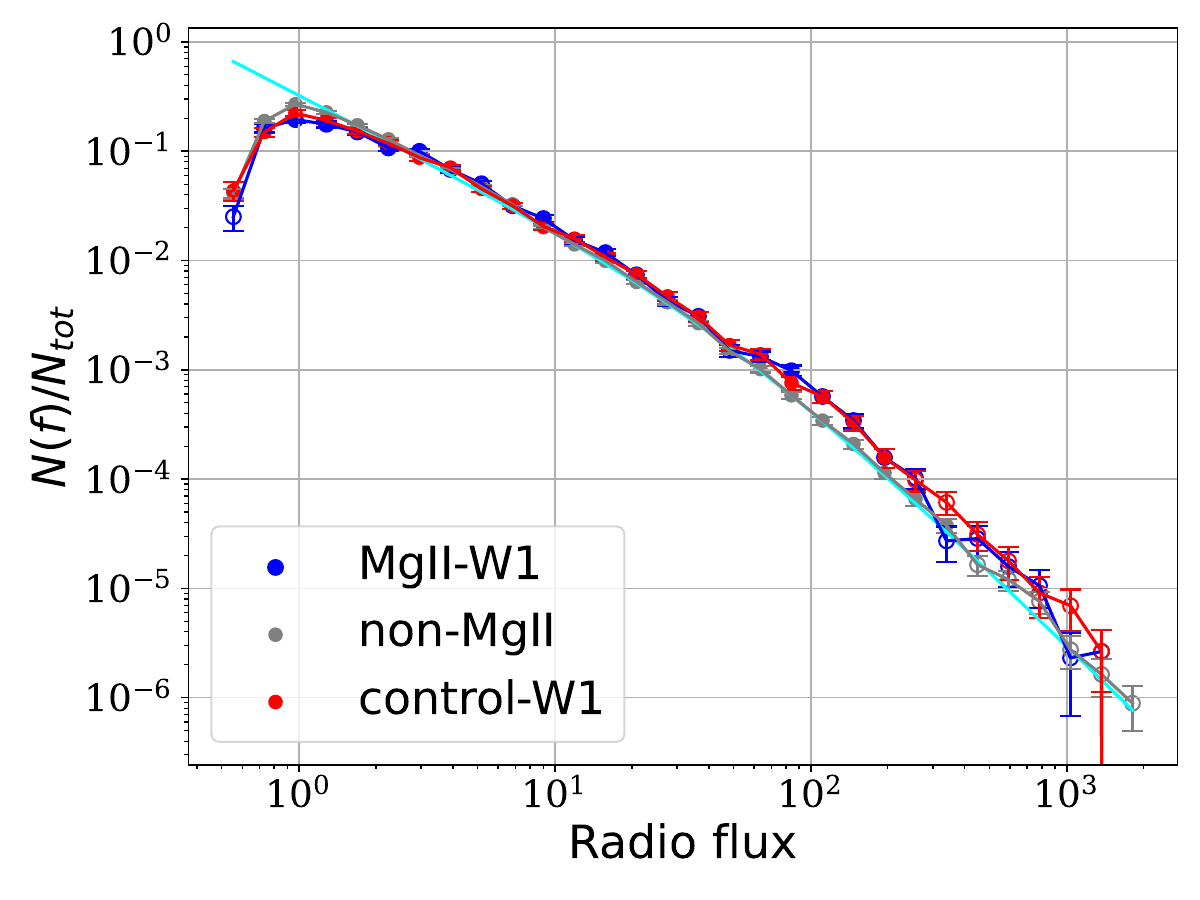}\\
\includegraphics[width=8cm]{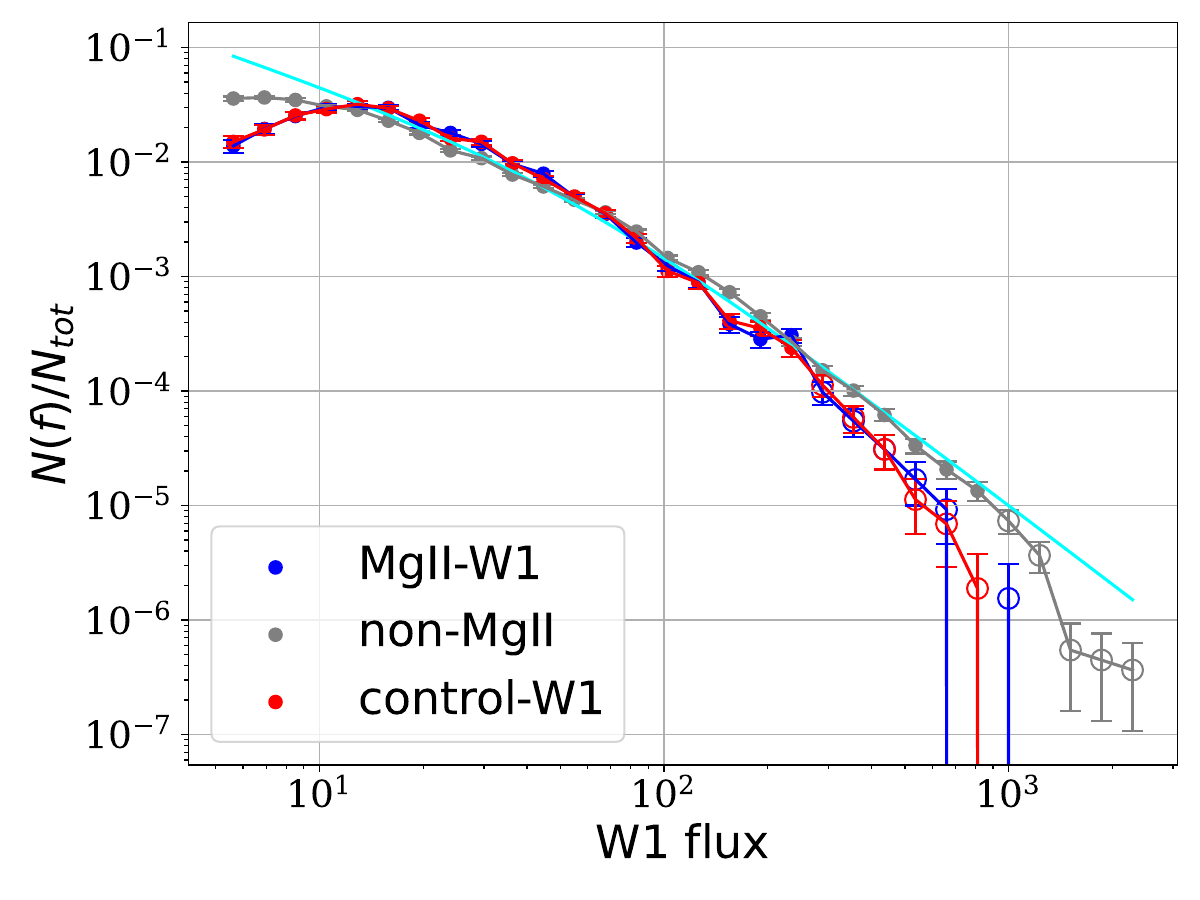}\\
\includegraphics[width=8cm]{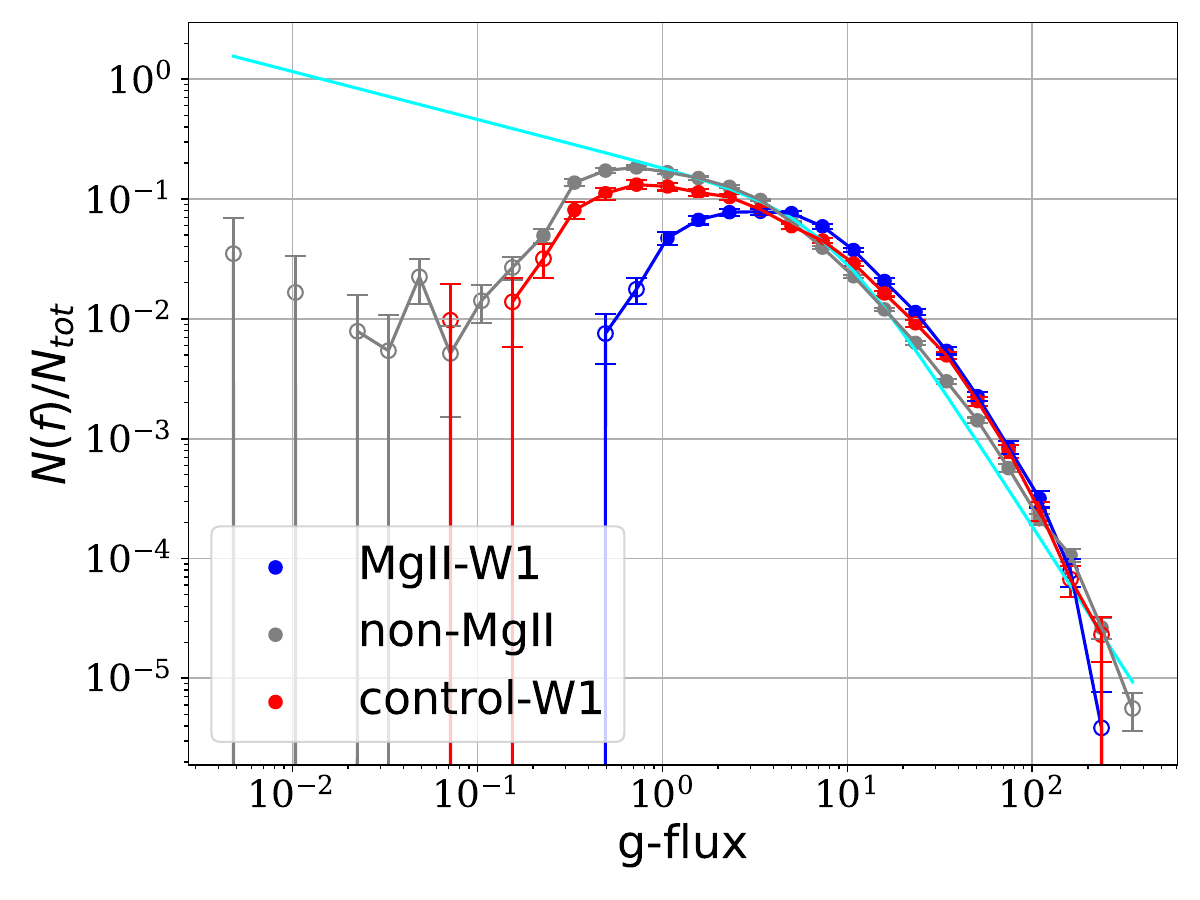}\\
\caption{
Histograms of radio flux (top), infrared W1 (middle), and optical $g-$band (bottom) for the three samples. Open circles indicate flux bins with fewer than 25 sources. 
The analytical curves (cyan) follow a broken power law with indices $a=1.2$, $b=2.3$ (top), $a=1$, $b=2.8$ (middle), $a=0.4$, $b=2.4$ (bottom) respectively. A modest excess at the bright end can be seen in radio band for both the MgII-W1 and control-W1 samples.}
\label{fig:radio-flux}
\end{figure}

Fig.\,\ref{fig:radio-flux} presents the source count histograms for the radio, W1 and $g$-band flux densities. For source counts less than $25$ in the flux bin, data are represented by open circles. The error bars present the relative Poisson noise. Across all bands and for all samples, the source counts exhibit a rapid decline with increasing flux. For reference, we show an analytical double power-law profile \citep[e.g.][]{2006AJ....131.2766R,2009MNRAS.399.1755C,2020MNRAS.495.3252S}
\be
\psi (f) \propto \dfrac{1 }{(f/f_b)^a + (f/f_b)^b},
\label{eq:analytical}
\ee
where $f_b$ is the broken flux, and $a$ and $b$ are the power index for the bright and faint side, respectively. We do not fit Eq.\,\ref{eq:analytical} to the data explicitly, but provide it for a qualitative guideline. An interesting trend is evident between the behaviour across bands. Although a double power-law is adopted for all three bands, the changing in slope in the radio is minimal, i.e. nearly a single power-law. As the frequency increases from radio to optical, the transition between the two power-law regimes becomes significant: the W1 and $g-$band distribution exhibit a flatter slope at the faint end and a steeper slope at the bright end. Such a changing of the flux distribution suggests that either the intrinsic luminosity functions differ significantly between these bands, or that band dependent propagation effects impart distinct signatures on the observed source counts.

The difference between the MgII sample and non-MgII samples is significant (Fig.\,\ref{fig:radio-flux}). While the radio flux distributions of the three samples (MgII-W1, control-W1, non-MgII) exhibit similar slopes across most of the flux range, the MgII-W1 and control-W1 samples show a notable excess at bright fluxes (>50 $\mu Jy$) relative to the non-MgII parent sample. 
This enhancement at the bright end resembles the characteristic signature of lensing magnification \citep[e.g.][]{2010MNRAS.406.2352L}. However, the similarity in behaviour between the MgII sample and its luminosity-matched controls suggests that if plasma lensing contributes, its imprint may be subtle or partially degenerate with other effects.
In the W1 band, the number counts at the bright end present steeper slopes for both the MgII and control samples, in contrast to the excess enhanced bright end behaviour observed in the radio band. This difference likely stems from the matching procedure used to construct the control samples. Specifically, the re-selection process excludes a large number of low-redshift sources, which dominate the population of bright infrared emitters (top middle panel in Fig.\,\ref{fig:z-2d-dist}). Since these low-redshift, high-flux W1 sources are largely removed from the control samples, and the MgII sample itself is intrinsically biased toward higher redshifts, the resulting W1 number counts are depleted at the bright end, leading to a steeper slope.
%The MgII sample excesses resample at all high flux range $\sim20-500$ in W1 band, and $\sim30-1000$ in W2 band. It has similar slope with the initial non-MgII sample at middle range of flux ($50-500$) and drops faster at bright end.
%

The optical flux distribution shows a mild evolution across the DESI bands; we therefore focus our analysis on the $g$-band. First, a clear distinction is evident between the MgII and non-MgII samples in both the two dimensional redshift-magnitude distribution (Fig.\,\ref{fig:z-2d-dist}) and the one dimensional flux histogram (Fig.\,\ref{fig:radio-flux}). The MgII sample contains only relatively bright sources (>0.5 nanomaggy), whereas the non-MgII sample extends to significantly fainter fluxes. 
%This deficit of faint MgII sources may arise from dust extinction in the absorbing systems.
Second, the source counts in all samples decline sharply toward the faint end, in addition to the turnover at the bright end. This steep drop likely reflects incompleteness of the sample near the optical flux limit of the survey. Despite these selection effects, the persistent offset between the MgII and non-MgII samples suggests that dust extinction and possibly plasma lensing in the radio band contribute jointly to the observed differences. 

\begin{figure}
\centering
\includegraphics[width=8cm]{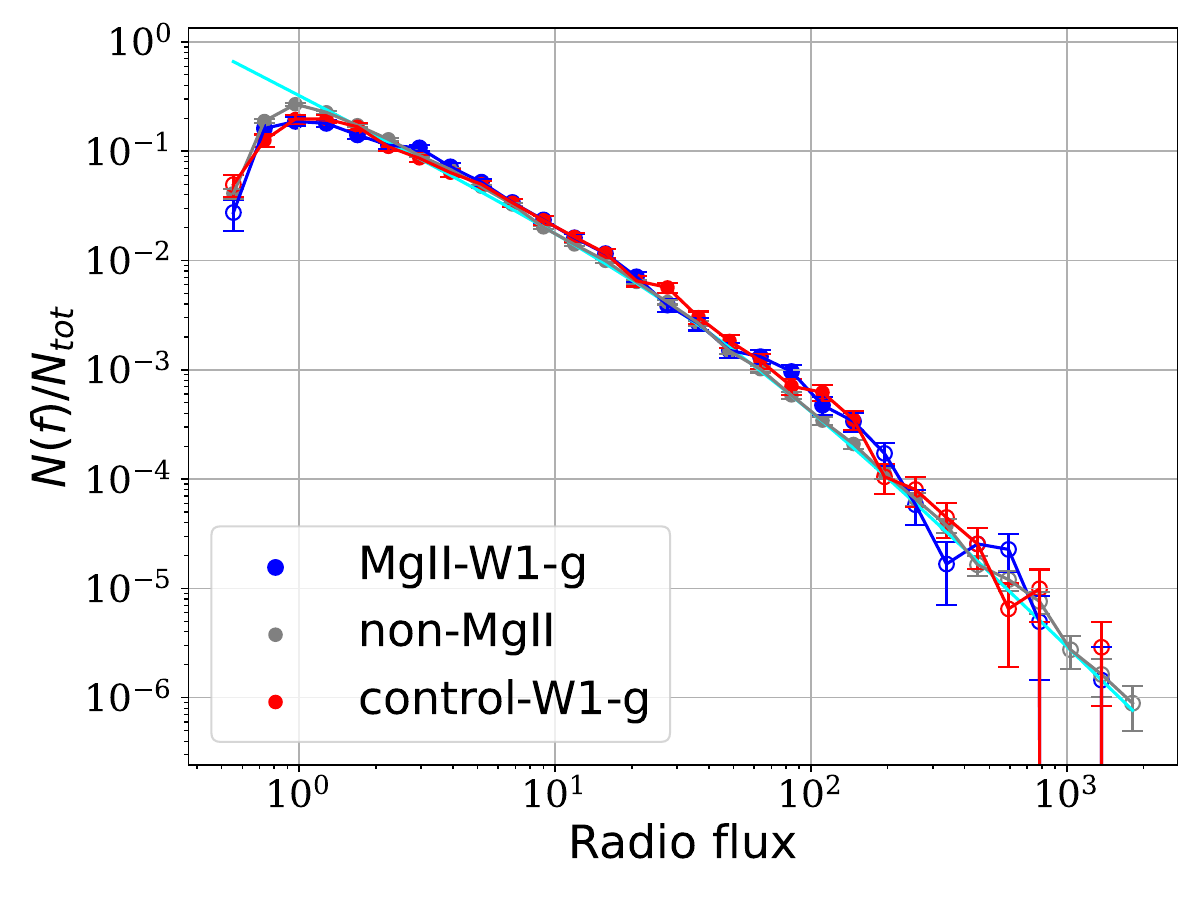}
\caption{Similar as Fig.\,\ref{fig:radio-flux} for only radio flux using control-W1-g sample. The bright-end excess is less pronounced compared to the MgII-W1 and control-W1 samples.}
\label{fig:flux-g-w1}
\end{figure}

For the control-W1-g sample, we present the number counts only in the radio band (Fig.\,\ref{fig:flux-g-w1}), as the control sample is by construction nearly identical to the MgII sample in both $g$-band and in W1 band luminosity distributions. In the radio band, the MgII sample and the control sample exhibit only a minor excess at bright fluxes relative to the non-MgII sample. It is noteworthy that the MgII sample exhibits a very slightly steeper decline at the faint end, a trend qualitatively consistent with plasma lensing predictions, though the difference is not statistically significant and lies within the uncertainties. To further test for potential dependencies on absorber strength, we divided the MgII sample based on the rest frame equivalent width of the MgII absorption \citep{2025arXiv250717866K}. However, no significant difference in radio number counts were found between subsamples with strong versus weak absorption systems.
%It is possible that the strong MgII absorption come from the Circumgalactic Medium (CGM) of the host galaxy, and has little lensing effects. 
%Another sample of sources with multiple absorption lines has been built. A slightly stronger excess has been found in the bright end of radio distribution.

\section{Conclusion and Discussion}
We have constructed a multi-wavelength QSO sample by cross matching optical photometry and spectroscopy from DESI DR1, infrared data from WISE, and radio observations from VLASS. The sample is then split according to the presence of MgII absorption lines, which indicates the foreground gas clumps along the line of sight.
The redshift distributions of the MgII sample and non-MgII sample show significant differences. To isolate the impact of intervening gas from selection effects, we constructed control samples by re-selecting sources from the non-MgII sample to match the MgII population in redshift and luminosity, either in the W1 band, or both in the optical $g$-band and W1 band simultaneously. The joint $g$–W1 control sample provides the most stringent null test in this work, effectively removing differences in intrinsic AGN power and optical selection, and leaving propagation effects as the primary remaining source of any residual radio excess.
Comparisons of number counts across radio, infrared, and optical bands reveal significant discrepancies between the MgII samples and the control samples. Notably, the MgII-W1 sample shows an excess at the bright end of the radio flux distribution, a signature that could arise from plasma lensing magnification or be influenced by dust extinction in absorbing systems. However, when using the most stringent control sample by matching jointly in redshift, $g$-band, and W1 luminosity, the radio excess becomes less pronounced. In the optical $g-$band, the MgII-W1 sample exhibits a steeper decline at the faint end compared to the control-W1 sample.
The radio excess and subsequent disappearance may suggests the existence of weak structures in intervening plasma clumps. However, since plasma lensing is strongly frequency dependent, its signature is expected to be weak at 3 GHz. Confirmation will therefore require low-frequency observations, such as \citep[e.g., LOFAR][]{2025A&A...695A..80S}. A direct test would be to compare the image properties of strongly lensed QSOs across different frequency bands, such as image positions, magnifications and arrival time differences.

Our analysis reveals distinct photometric behaviours between QSOs with and without MgII absorption. 
In all three bands (radio, infrared W1, optical $g$), the two dimensional distributions of redshift and flux (magnitude) show clear differences between the MgII and the control samples.
While we found no significant correlation between radio flux and colour index ($g-W1$), a weak positive correlation is present between radio flux and both W1 and $g$-band flux in the MgII and control samples. 
%This correlation suggests a possible scenario where dust re-emission contributes to the emission in the middle infrared bands. 
Moreover, the source count distributions evolve systematically with wavelength: they are well described by a single power law in the radio band but transition to a distinct double power-law form in the optical $g$-band.
A clear anti-correlation between redshift and infrared flux is observed across all samples, which we attribute to selection effects, e.g., the infrared colour cuts employed in the DESI target selection. No such feature is evident in the radio flux distribution.

Our samples span a broad redshift range, with a distribution peaking at a relatively high redshift, e.g., $z\sim2$. The source counts presented here are based on observed flux without applying a $K-$correction. Consequently, a given observed flux originates from different rest frame wavelengths for sources at different redshifts. This introduces a spectral mixing effect, meaning the observed distributions can differ even in the absence of intrinsic cosmic evolution. This effect may dilute frequency dependent propagation effects, such as those expected from plasma lensing.
For sources at high-redshift, e.g. $z>2$, the likelihood of light from higher redshift sources interacting with intervening gas clumps is increased. These include not only dense gas responsible for detectable absorption lines but also more diffuse components, such as the circumgalactic medium (CGM) of foreground galaxies \citep[e.g.,][]{2017ApJ...850..156L,2021ApJ...923...56D}. Numerical simulations show that the sight-lines to such sources often intersect the CGM of multiple foreground galaxies \citep[e.g.][]{2025ApJS..277...43M}. As a result, plasma lensing becomes a non-negligible effect, particular at high redshift in low radio bands. Therefore, our multi-wavelength analysis serves a dual purpose: not only we can study the properties of background sources, but this also opens a new avenue for statistically probing the properties of diffuse cosmic plasma structures along the line of sight.

At the current stage, the data do not permit a definitive conclusion regarding the presence of plasma lensing. While our results do not rule out plasma lensing entirely, they suggest that it is unlikely to be the dominant effect shaping the observed radio number counts in this sample. The observed signatures are equally consistent with dust extinction and subsequent re-emission in foreground absorbing systems. Nevertheless, several aspects may influence our results and need further studies. 

\begin{itemize}
\item 
The incompleteness of the MgII sample at low redshifts may introduce contamination, potentially weakening the observed differences between the two samples.
\item 
The local environment of the QSO, particularly its host galaxy and jet, may contribute to the observed signal. Gas clumps associated with these structures can produce MgII absorption \citep[e.g.][]{2025ApJ...980..159C}. But if the gas clumps lie close to the background source, their lensing efficiency is minimal. 
\item 
The fibre size of DESI ($3''$) integrates light over a finite angular region, allowing for a small transverse offset between the gas clumps and the QSO sight line. Depending on the relative position, plasma lensing can produce either magnification or de-magnification \citep[e.g.][]{2022MNRAS.510..197E}, leading to a complex signal in ensemble statistics.
\item 
Constructing a truly representative control sample remains challenging. Matching in both redshift and multi-wavelength flux inevitably discards a significant fraction of sources, predominantly at high redshift (as seen in the middle and right panels of Fig.\,\ref{fig:z-dist}). The physical origin of this residual mismatch is unclear; it may reflect intrinsic differences in the populations, or limitations in current selection methods. 
\end{itemize}

\section*{Acknowledgements}
We thank the referee for his/her constructive and enlightening suggestions and comments on the manuscript.
This research used data obtained with the Dark Energy Spectroscopic Instrument (DESI). DESI construction and operations is managed by the Lawrence Berkeley National Laboratory. This material is based upon work supported by the U.S. Department of Energy, Office of Science, Office of High-Energy Physics, under Contract No. DE–AC02–05CH11231, and by the National Energy Research Scientific Computing Center, a DOE Office of Science User Facility under the same contract. Additional support for DESI was provided by the U.S. National Science Foundation (NSF), Division of Astronomical Sciences under Contract No. AST-0950945 to the NSF’s National Optical-Infrared Astronomy Research Laboratory; the Science and Technology Facilities Council of the United Kingdom; the Gordon and Betty Moore Foundation; the Heising-Simons Foundation; the French Alternative Energies and Atomic Energy Commission (CEA); the National Council of Humanities, Science and Technology of Mexico (CONAHCYT); the Ministry of Science and Innovation of Spain (MICINN), and by the DESI Member Institutions: www.desi.lbl.gov/collaborating-institutions. The DESI collaboration is honored to be permitted to conduct scientific research on I’oligam Du’ag (Kitt Peak), a mountain with particular significance to the Tohono O’odham Nation. Any opinions, findings, and conclusions or recommendations expressed in this material are those of the author(s) and do not necessarily reflect the views of the U.S. National Science Foundation, the U.S. Department of Energy, or any of the listed funding agencies.
VLASS: The National Radio Astronomy Observatory is a facility of the National Science Foundation operated under cooperative agreement by Associated Universities, Inc. CIRADA is funded by a grant from the Canada Foundation for Innovation 2017 Innovation Fund (Project 35999), as well as by the Provinces of Ontario, British Columbia, Alberta, Manitoba and Quebec.

%%%%%%%%%%%%%%%%%%%% REFERENCES %%%%%%%%%%%%%%%%%%
\section*{Data Availability}
The data underlying this article will be shared on reasonable request to the corresponding author.
%%%%%%%%%%%%%%%%%%%%
\bibliographystyle{mnras}
\bibliography{reference}

%%%%%
\appendix
\section{More comparisons on flux}
\begin{figure}
\centering
\includegraphics[width=8cm]{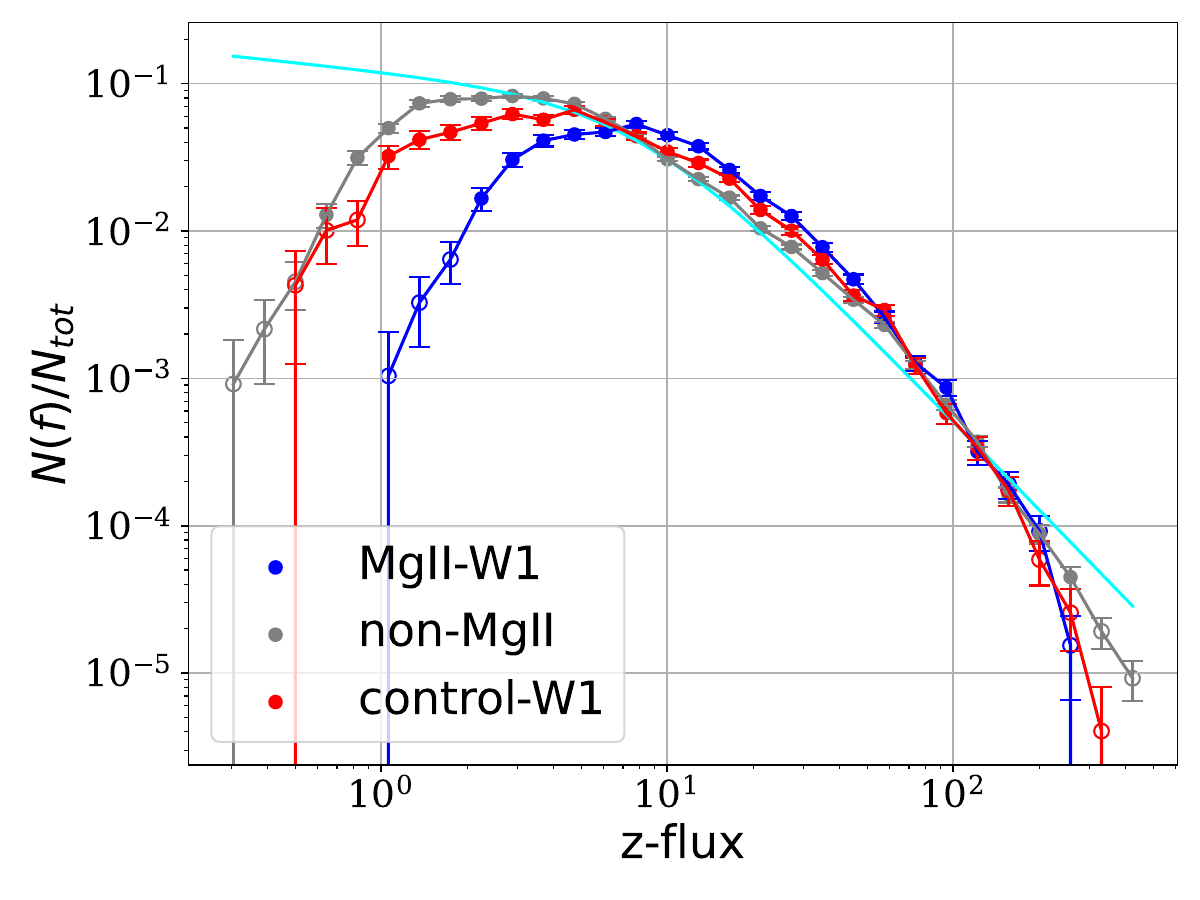}
\caption{Similar as Fig.\,\ref{fig:radio-flux} for z-band flux.}
\label{fig:z-flux}
\end{figure}

\begin{figure}
\centering
\includegraphics[width=7cm]{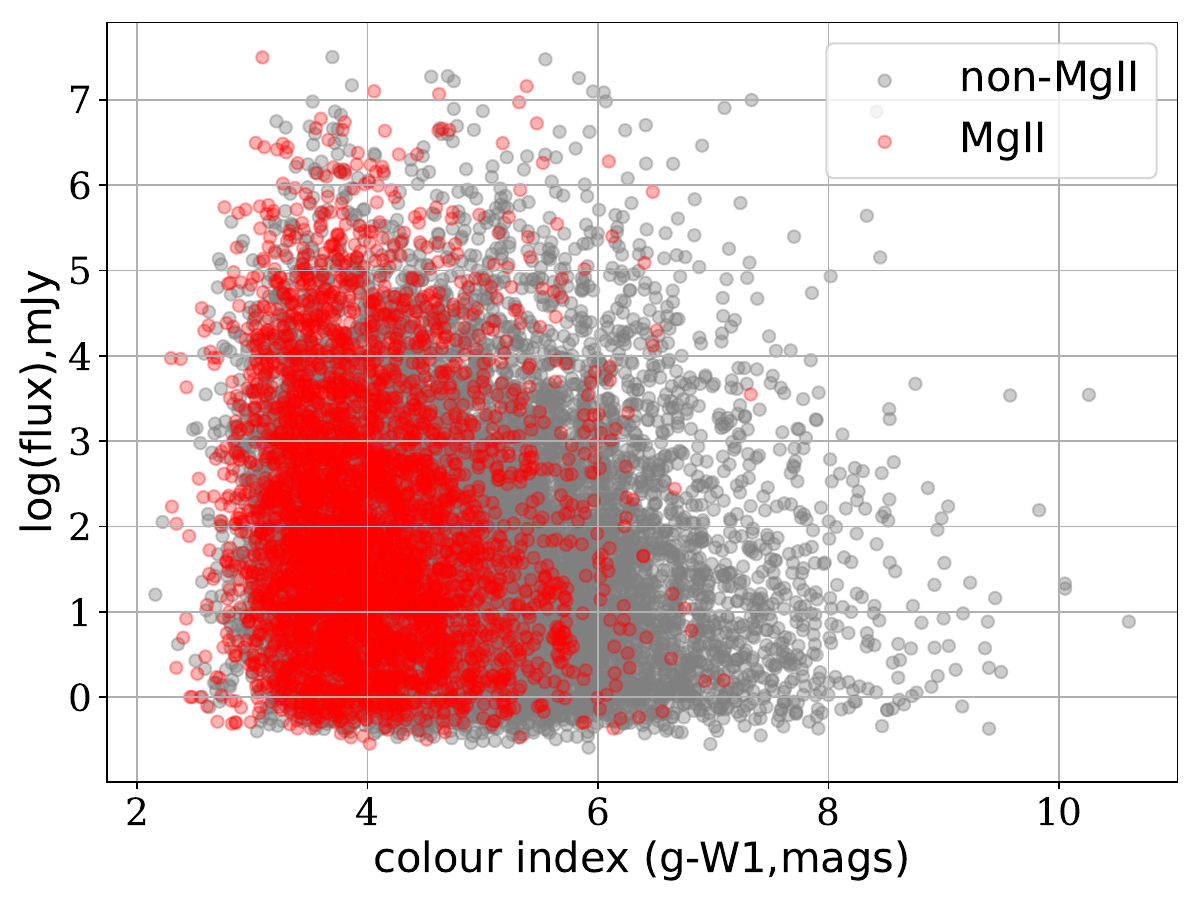}
\caption{Source distribution in the (g-W1) colour index radio flux plane. Red (grey) dots show sources with (without) MgII absorption.}
\label{fig:color2fr}
\end{figure}

In this section, we provide supported distributions of flux in different samples. In Fig.\,\ref{fig:z-flux}, the number counts in z-band flux is presented. In Fig.\,\ref{fig:color2fr}, the sources is presented on the plane of colour index ($g-W1$) and radio flux. No significant correlation between $g-W1$ and radio flux is found, suggesting that dust re-emission alone may not fully account for the radio excess at the bright end.

\begin{figure*}
\centering
\includegraphics[width=5.5cm]{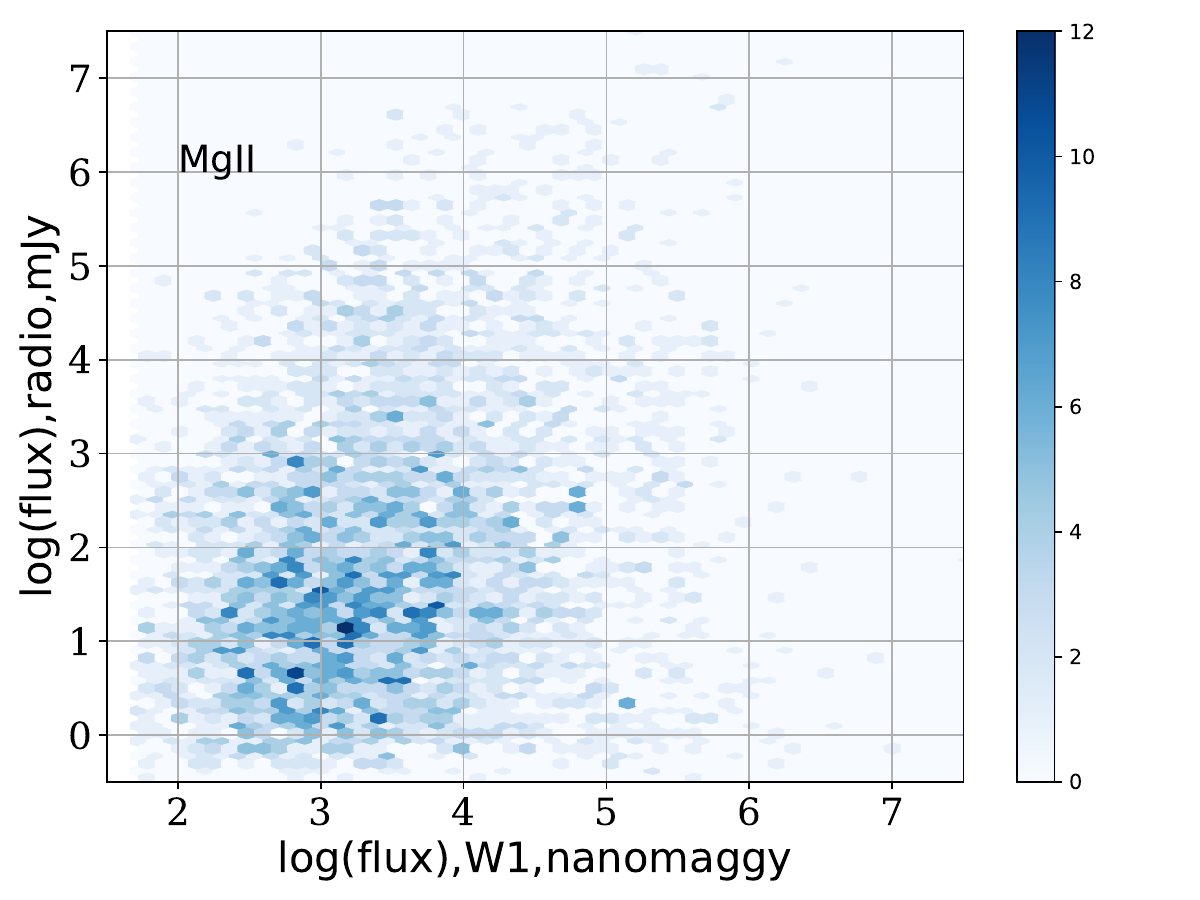}
\includegraphics[width=5.5cm]{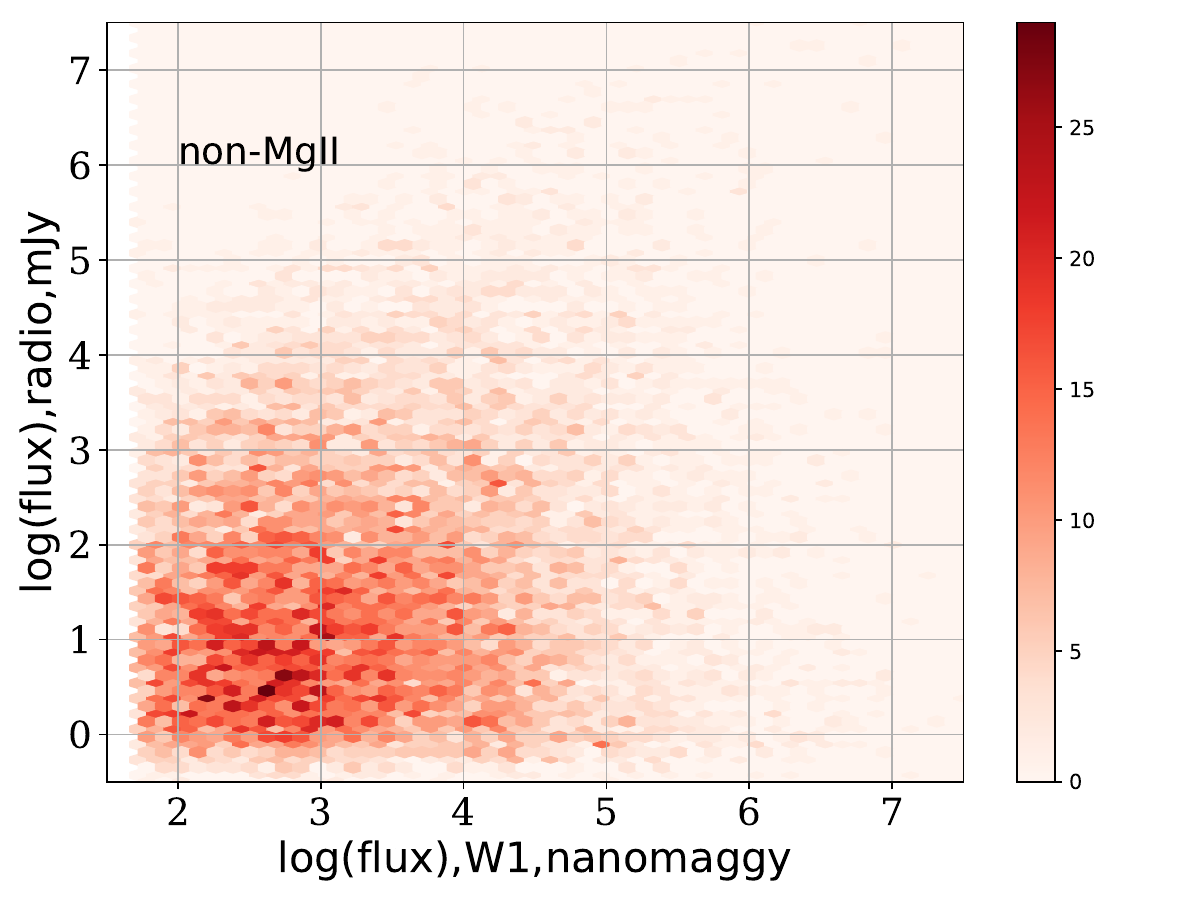}
\includegraphics[width=5.5cm]{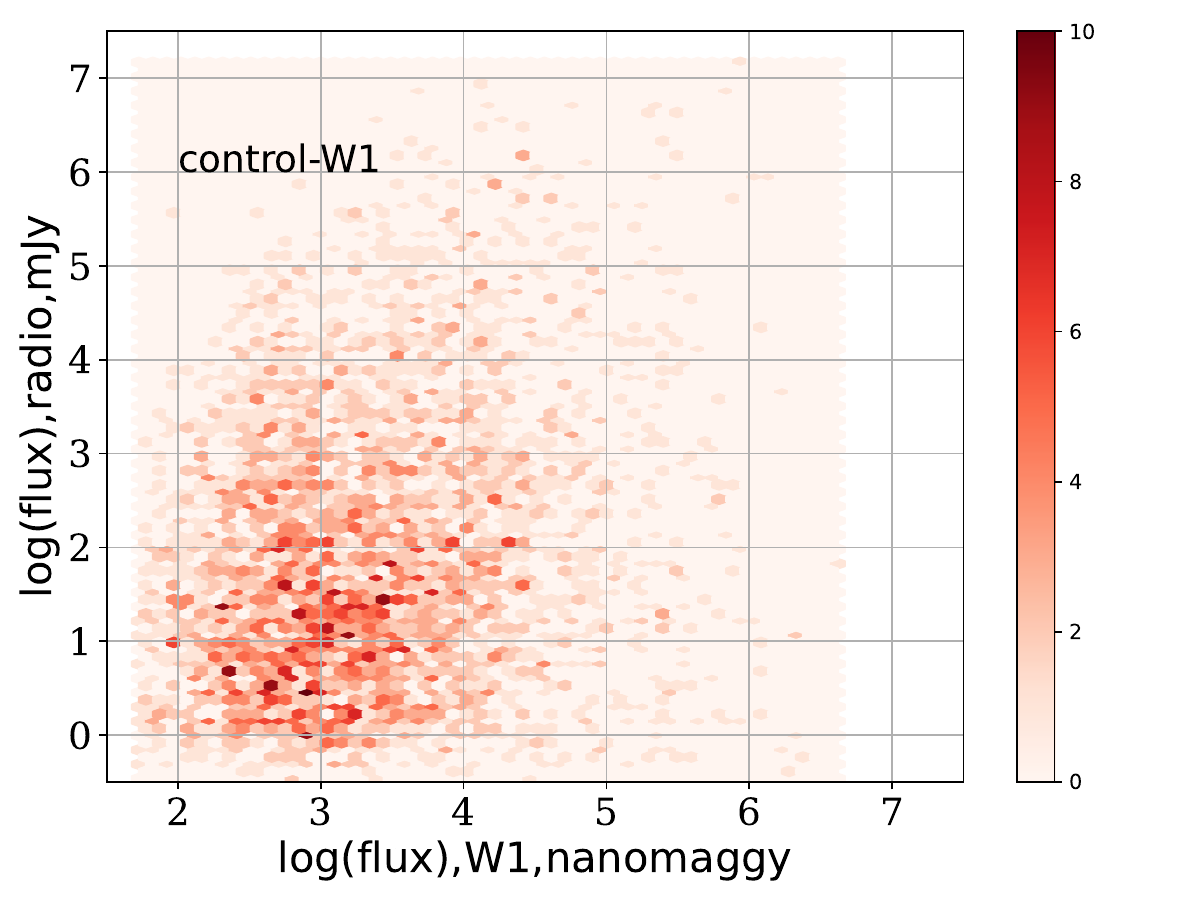}
\caption{Distribution of QSOs in the infrared (W1) versus radio flux density plane. }
\label{fig:corr-w1-radio}
\end{figure*}

\begin{figure*}
\centering
\includegraphics[width=5.5cm]{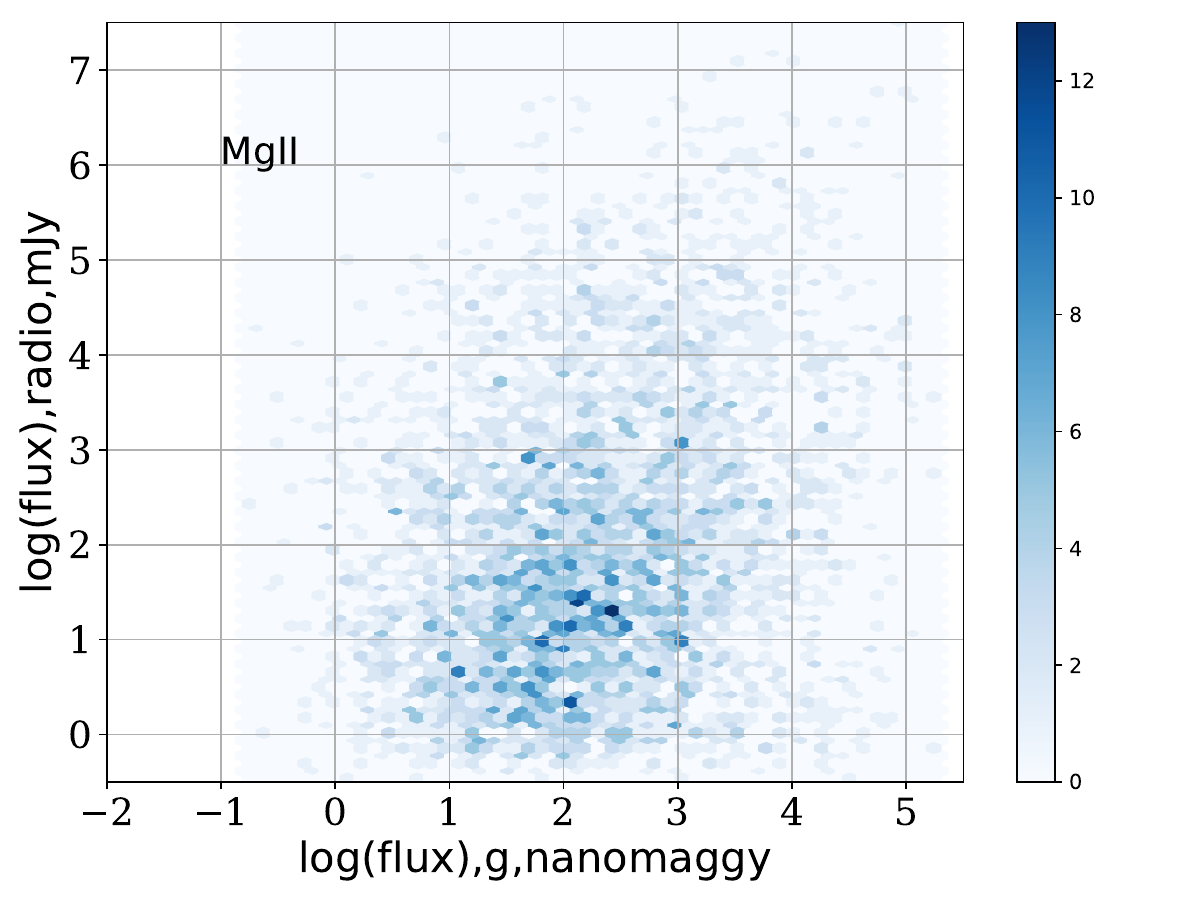}
\includegraphics[width=5.5cm]{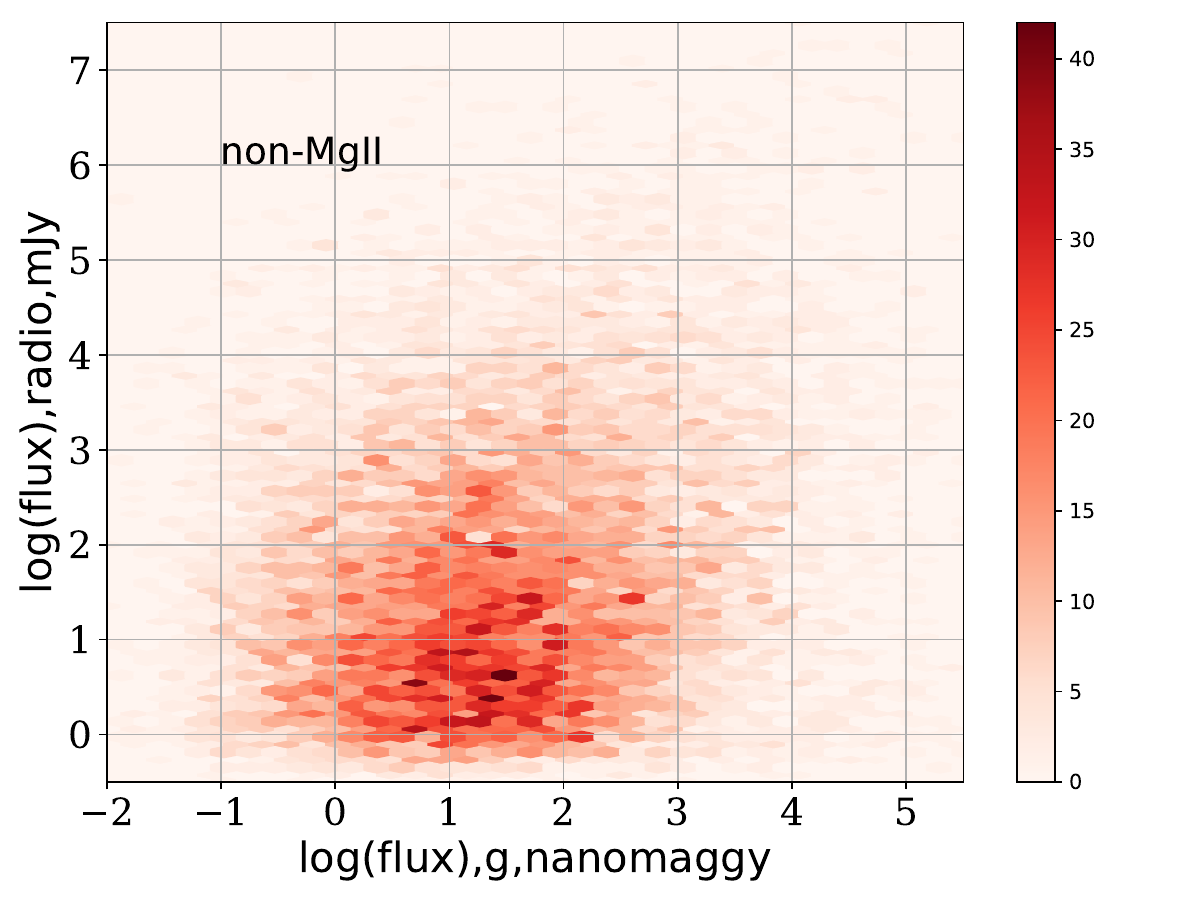}
\includegraphics[width=5.5cm]{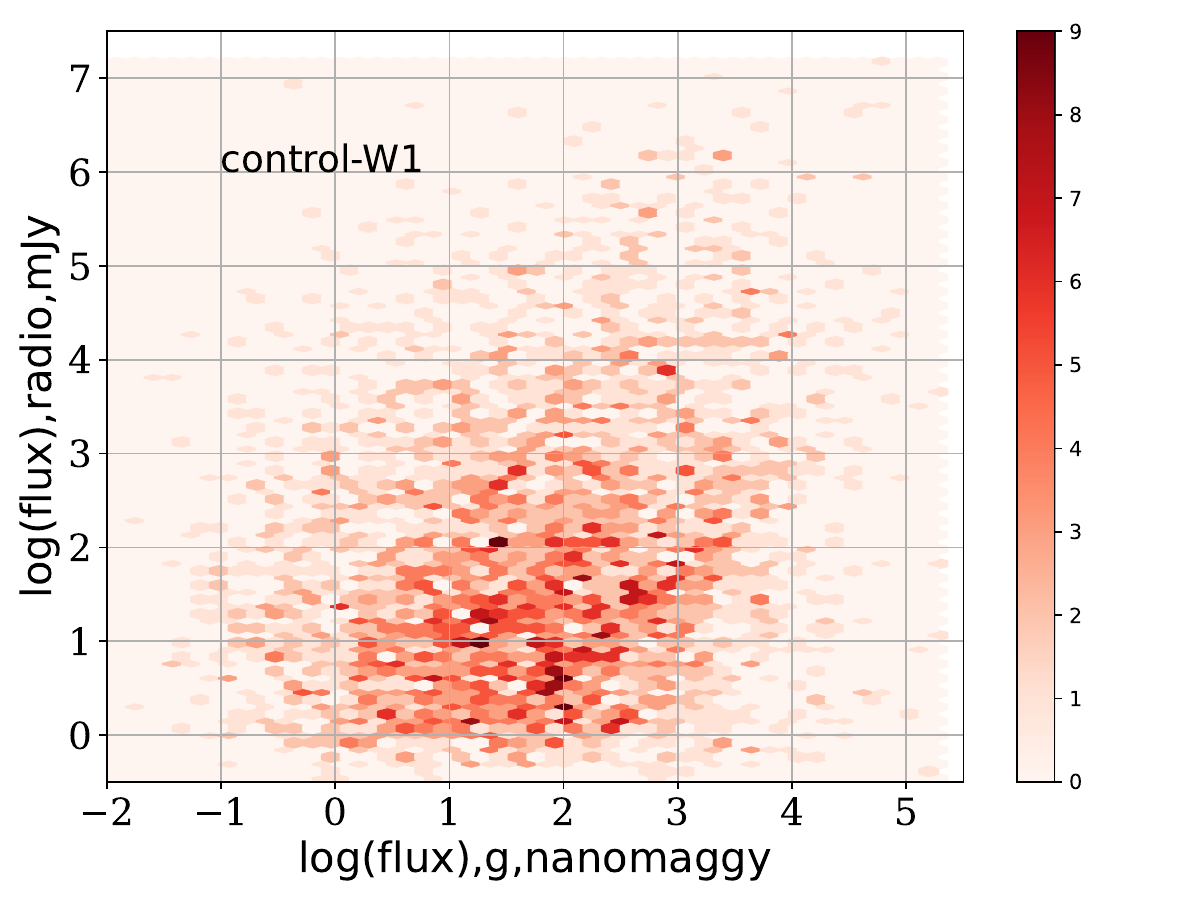}
\caption{Same as Fig.\,\ref{fig:corr-w1-radio} but for the optical ($g$-band) versus radio flux density. }
\label{fig:corr-g-radio}
\end{figure*}

In Figs.\,\ref{fig:corr-w1-radio} and \ref{fig:corr-g-radio}, we present the distribution of sources on radio flux with W1 ($g$-band) flux. Both the MgII and control-W1 samples exhibit a weak correlation between the radio flux and W1 (or $g-$band) fluxes, suggesting that the observed radio excess may not originate solely from plasma lensing.

%

% Don't change these lines
\bsp	% typesetting comment
\label{lastpage}

\end{document}